\documentclass[aps, pra,superscriptaddress,twocolumn,a4paper]{revtex4-1}
\usepackage{bm}
\usepackage{appendix}
\usepackage{graphicx}
\usepackage{amsmath}
\usepackage{amssymb}%
\usepackage{color}
\usepackage{lipsum}
\usepackage[colorlinks=true,citecolor=blue,urlcolor=blue]{hyperref}
\begin{document}
\title{Ground-state phase diagram and excitation spectrum of a Bose-Einstein condensate with spin-orbital-angular-momentum coupling}

\author{Ke-Ji Chen}
\affiliation{Department of Physics and State Key Laboratory of Low-Dimensional
Quantum Physics, Tsinghua University, Beijing 100084, China}

\author{Fan Wu}
\affiliation{Department of Physics and State Key Laboratory of Low-Dimensional
Quantum Physics, Tsinghua University, Beijing 100084, China}

\author{Jianshen Hu}
\affiliation{Department of Physics and State Key Laboratory of Low-Dimensional
Quantum Physics, Tsinghua University, Beijing 100084, China}

\author{Lianyi He}
\email{lianyi@mail.tsinghua.edu.cn}
\affiliation{Department of Physics and State Key Laboratory of Low-Dimensional
Quantum Physics, Tsinghua University, Beijing 100084, China}

\date{\today}
\begin{abstract}
We investigate the ground-state phase diagram and excitation spectrum of an interacting spinor Bose-Einstein condensate with spin-orbital-angular-momentum (SOAM) coupling realized in recent experiments by introducing atomic Raman transition with a pair of copropagating Laguerre-Gaussian laser beams that carry different orbital angular momenta (OAM) [Chen \emph{et al}., Phys. Rev. Lett. {\bf 121}, 113204 (2018) and Zhang \emph{et al.}, Phys. Rev. Lett. {\bf 122}, 110402 (2019)]. Because of the ground-state degeneracy of the single-particle Hamiltonian at vanishing detuning, several angular-stripe phases, which are superposition of states with different angular quantum numbers, appear in the phase diagram. However,  these phases normally exist at small detuning, which makes them hard to be probed in experiments. We show that for a large OAM difference of the laser beams, an asymmetric kind of angular-stripe phase can exist even at large detuning. The excitation spectra in different phases exhibit distinct features: In the angular-stripe phase there exist two gapless bands corresponding to the broken U$(1)$ and rotational symmetries, while in the half-skyrmion phase the gapless band exhibits a roton-like structure. Our predictions of the angular-stripe phases and the low-energy excitations can be examined in recently realized BECs with SOAM coupling.
\end{abstract}

\maketitle
\section{introduction}\label{introduction}
Spin-orbit coupling (SOC), i.e., coupling between a particle's spin and orbital motion, gives rise to many fascinating phenomena, such as quantum spin Hall effects ~\cite{kato04, kane-05, zhang-07, zhang-06}, topological insulators, and topological superconductors ~\cite{kane-10, zhang-11}. Ultracold atoms with the high controllability of degrees of freedom provide an ideal platform to simulate and study those exotic physical effects by creating  synthetic gauge potentials ~\cite{dalibard-11, spielman-13, zhai-15}.   A prominent example  is  the so-called  spin-linear-momentum  (SLM) coupling, which is achieved by introducing Raman transitions among the internal states of atoms by two counterpropagating laser fields ~\cite{lin-11, zhang-12, zwierlein-12}. A variety of exotic quantum states have been observed in quantum gases with SLM coupling ~\cite{dalibard-11, spielman-13,zhai-15, lin-11, zhang-12, zwierlein-12, shuai-16, engels-17, ketterle-17}. 

In addition to the SLM coupling, another fundamental type of SOC,  the spin-orbital-angular-momentum (SOAM) coupling, was theoretically proposed ~\cite{pu-15, sun-15} and recently realized in spin-1/2 and spin-1 $^{87}$Rb  Bose gases~\cite{jiang-19, lin-18} by introducing atomic Raman transition with a pair of copropagating Laguerre-Gaussian (LG) optical fields that carry different orbital angular momentum (OAM).  Different from SLM coupling, SOAM coupling preserves the rotational symmetry and leads to a discrete spectrum. Thus a spinor Bose-Einstein condensate (BEC) subjected to SOAM coupling may exhibit distinct properties compared to SLM coupling.  Previous theoretical studies have predicted several exotic phases, such as the vortex-antivortex pair phase, half-skyrmion phase, and angular-stripe phase ~\cite{qu-15, chen-16,hu-15, hou-17, hu-19}. 

\begin{figure*}[t]
\begin{center}
\includegraphics[width=0.7\textwidth]{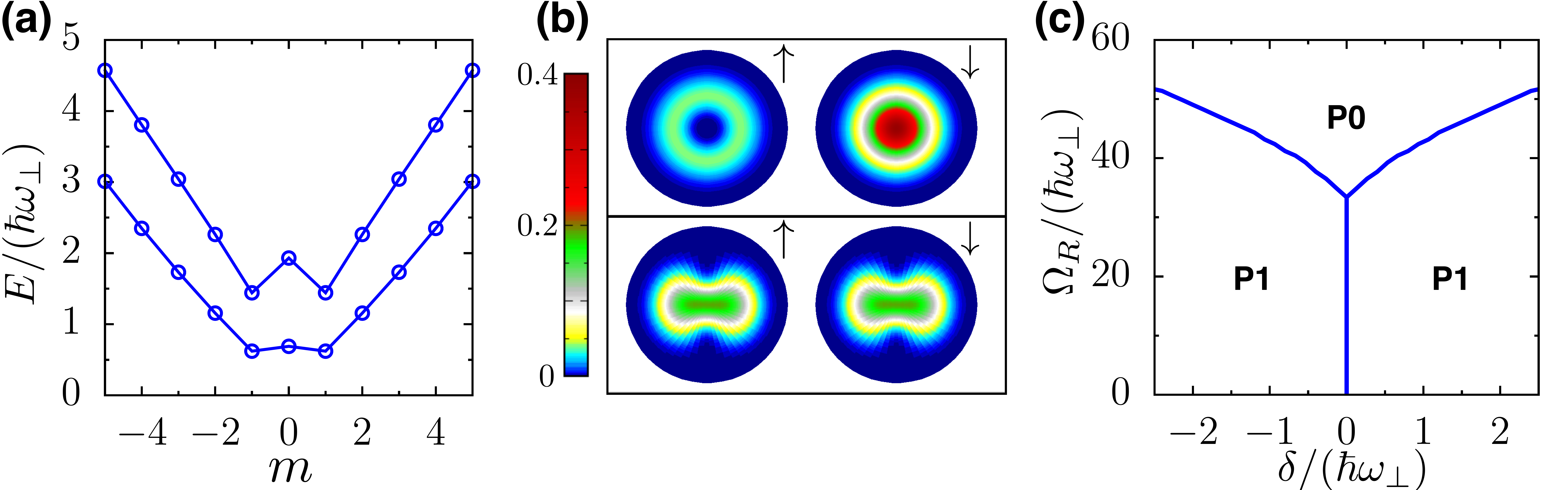}
\caption{(a) Energy spectrum of the single-particle Hamiltonian $\hat{\cal H}_s$ with $\delta=0$ and $\Omega_R/(\hbar \omega_\perp)=30$ . (b) Density profiles of a noninteracting Bose gas with SOAM coupling at $\delta=0$ and $\Omega_R/(\hbar \omega_\perp)=30$: $\phi=0$ (top) and $ \phi=\pi/4, \gamma=0$ (bottom).  (c) Phase diagram of a noninteracting Bose gas with SOAM coupling. Here P1 denotes the phase where the bosons condense in the $m=l$ or $m=-l$ state, and in the P0 phase the bosons condense in the $m=0$ state. In these plots, we consider the case $l=1$. The trapping frequency is $\omega_{\bot}=2\pi\times 10$Hz and the waist of  LG laser is $w=25\mu m$. }
\label{noninteraction}
\end{center}
\end{figure*}

In this paper, we present a systematic theoretical study of the ground-state phase diagram and excitation spectrum of a two-dimensional (2D) spinor BEC with SOAM coupling realized in recent experiments ~\cite{jiang-19, lin-18}.  An important feature of SOAM coupling is that it leads to ground-state degeneracy of the single-particle spectrum (Fig. \ref{noninteraction}). Similar ground-state degeneracy was also found in spinor BEC with SLM coupling, leading to the exotic stripe phase ~\cite{wang-10, wu-11, ho-11, li-12, li-13, liao-18}.  Here we expect that the degeneracy leads to the angular-stripe phase, where the many-body ground state is a superposition of different states with definite angular quantum numbers. However, in a realistic system, detuning and Raman coupling are unavoidable. They lift the degeneracy and hence kill the angular-stripe phase in a noninteracting spinor BEC.

The intra- and interspecies interactions are also unavoidable, leading to rich ground-state phases. At small interspecies interaction, there exists a room for the angular-stripe phase (Fig. \ref{interacting-phase}). For the OAM difference $l=1$ of the LG laser beams used in present experiments~\cite{jiang-19, lin-18}, the largest detuning for the angular-stripe phase is rather small and hence it is hard to be probed in experiments. However, for larger OAM difference $l$, we find an unconventional angular-stripe phase which exists even at large detuning [see Fig.  \ref{interacting-phase}(b) for $l=2$]. This may provide a new possibility to observe the angular-stripe state. The excitation spectrum of a spinor BEC with SOAM coupling is also studied. In the angular-stripe phase, the excitation spectrum exhibits two gapless bands, corresponding to the broken U$(1)$ symmetry and rotational symmetry.  In the half-skyrmion phase,  only one gapless band appears and exhibits a roton-like structure. These distinct features of the excitation spectra in different phases can also be probed in future experiments.

The paper is organized as follows.  In Sec. \ref{model}, we establish the model, analyze the single-particle spectrum, and discuss the origin of the angular-stripe phase.   In Sec. \ref{phase-diagram},  we investigate the ground-state phase diagram of an interacting BEC with SOAM coupling.  In Sec. \ref{spectrum} the low energy excitations of different phases are studied. We summarize in Sec. \ref{summary}. 

\section{The model}\label{model}
We consider a 2D spin-1/2 Bose gas with SOAM coupling confined in a 2D harmonic potential. In the rotating frame, the system can be described by the grand-canonical Hamiltonian $H=H_0+H_{\rm int}$, where the noninteracting part reads
\begin{eqnarray}
\begin{split}
H_0 =\int d{\bf r} \psi^{\dag}\left(\hat{\cal H}_s-\mu \right)\psi,
\end{split}
\label{Hamiltonian-0}
\end{eqnarray}
where $\psi=(\psi_\uparrow,\ \psi_\downarrow)^{\rm T}$ denotes the spin-1/2 boson field with mass $M$ and chemical potential $\mu$.  In the polar coordinates ${\bf r}=(r,\theta)$, the single-particle Hamiltonian is given by ~\cite{jiang-19}
\begin{eqnarray}
\hat{\cal H}_s&=&-\frac{\hbar^2}{2Mr}\frac{\partial}{\partial r}\left(r\frac{\partial}{\partial r}\right)+\frac{(L_z-l\hbar\sigma_z)^2}{2Mr^2}\nonumber\\
&&+\ \Omega(r)\sigma_x+\frac{\delta}{2}\sigma_z+V_{\text{ext}}({\bf r}),
\label{Hamiltonian-s}
\end{eqnarray}
where $L_{z}=-i\hbar \partial /\partial \theta$ is the OAM along the $z$ axis, which couples to the atom spin via the SOAM coupling term $\sim L_z\sigma_z$. The spatial-dependent coupling of the laser beams is given by 
\begin{eqnarray}
\Omega(r)= \Omega_{R}\left(\frac{r}{w}\right)^{|l_{1}|+|l_{2}|} e^{-2\frac{r^2}{w^2}},
\end{eqnarray}
with the Rabi frequency $\Omega_{R}$, waist $w$, and detuning $\delta$.  $l_1$ and $l_2$  are winding numbers of a pair copropagating LG laser beams and $l=(l_1-l_2)/2>0$. Without loss of generality, we consider the case $l_1=0$ and $l_2=-2l$ as experimentally realized ~\cite{jiang-19, explain01}. The external confinement potential is given by $V_{\text{ext}}({\bf r})=M\omega_\perp^2r^2/2$.

Before discussing the effects of interactions, an analysis of single-particle Hamiltonian will be helpful to understand the origin of stripe phase. In the absence of interactions,  the energy spectrum and eigenstates can be obtained straightforwardly by solving the following eigenvalue equation 
\begin{eqnarray}
\hat{\cal H}_s\psi_\eta({\bf r})=E_\eta\psi_\eta({\bf r}).
\end{eqnarray}
Because of the rotational symmetry, i.e., $[\hat{\cal H}_s,L_z]=0$, we can separate the angular part by writing  $\psi_\eta({\bf r})=\varphi_n(r)\Theta_m(\theta)$, where $\Theta_m(\theta)=e^{i m\theta}/\sqrt{2\pi}$.  The radial part $\varphi_n(r)$ and the energy spectrum $E_{n}(m)$ can be determined numerically. For vanishing detuning ($\delta=0$) and small Rabi frequency 
$\Omega_R$, the ground state generally exists two-fold degeneracy for $m=l$ and $m=-l$, as shown in Fig.\ref{noninteraction} (a). The ground-state degeneracy  gives  rise to the possibility of an angular-stripe state, since the superposition state
\begin{eqnarray}
\psi^{s}_{G}({\bf r}) =  \cos\phi\  \psi^{+l}_{G}({\bf r})+\sin\phi\  e^{i \gamma}  \psi^{-l}_{G}({\bf r}),
\end{eqnarray}
with $\psi^{\pm l}_{G}({\bf r})=\varphi_G(r)\Theta_{\pm l}(\theta)$ being the two degenerate ground states,  also has the minimal energy. Here $\gamma$ is an arbitrary real number and $\phi \in [0,\frac{\pi}{2}] $. The superposition state with $\phi \neq 0 $ and  $\phi \neq\frac{\pi}{2}$ is the so-called angular-stripe state, of which the density profile modulates along the angular direction, as shown in Fig.~\ref{noninteraction}(b).  Therefore, for vanishing detuning ($\delta=0$) and small Rabi frequency $\Omega_R$, the superposition state with arbitrary $\phi$ and $\gamma$ can be the ground state of a noninteracting Bose gas with SOAM coupling.

In Fig.~\ref{noninteraction}(c), we show the phase diagram of a noninteracting Bose gas with SOAM coupling. In the presence of detuning ($\delta\neq0$),  or at large Rabi frequency $\Omega_R$, ground-state degeneracy is broken and the angular-stripe state is no longer the ground state.  So far we have not yet considered the interaction effect.  The interaction Hamiltonian $H_{\rm int}$ is given by
\begin{eqnarray}
\begin{split}
H_{\rm int}=\int  d{\bf r}\left[\frac{g}{2}(\hat{n}_\uparrow^2+\hat{n}_\downarrow^2)+g_{\uparrow \downarrow}\hat{n}_\uparrow \hat{n}_\downarrow\right],
\end{split}
\label{Hamiltonian-int}
\end{eqnarray}
where $\hat{n}_\sigma=\psi_\sigma^\dagger\psi_\sigma^{\phantom{\dag}}$ $(\sigma=\uparrow,\downarrow)$ is the density operator,  and $g$ and $g_{\uparrow \downarrow}$ denote the contact intra- and interspecies interactions, respectively. 
The presence of  interactions make the ground state of the system subtle. We note that the interaction Hamiltonian can be written as a more illustrating form,
\begin{eqnarray}
H_{\text{int}}  =  \int d {\bf r} [c_{0}\left( \hat{n}_{\uparrow}+\hat{n}_{\downarrow}\right)^2+c_{1}\left(\hat{n}_{\uparrow}-\hat{n}_{\downarrow}\right)^2], 
\end{eqnarray}
where  $c_{0}=(g+g_{\uparrow \downarrow})/4$ and $c_{1}=(g-g_{\uparrow \downarrow})/4$. When $g<g_{\uparrow \downarrow}$, the system tends to be polarized to decrease the interaction energy, which disfavors the stripe phase.  However,  for $g>g_{\uparrow \downarrow}$,  the density profiles of two spin components tend to be balanced, which makes the stripe phase more favorable. As we will show below, for $g>g_{\uparrow \downarrow}$, the interaction effect will make the angular-stripe state more stable against the detuning.  The system thus has a rich phase structure due to the competition among the detuning, Raman coupling, and the interaction.

\section{Ground-state phase diagram}\label{phase-diagram}
The system can be conveniently studied by using the imaginary-time functional path integral approach, where the action is given by 
\begin{eqnarray}
{\cal S}  =  \int_0^\beta d \tau \left[\psi^{\dagger}\partial_\tau\psi+H(\psi,\psi^\dagger)\right],
\end{eqnarray}
with $\tau$ being the imaginary time and $\beta=1/(k_{\rm B}T)$. In the BEC state,  the field $\psi_{\sigma}(\tau, {\bf r})$ can be decomposed as  
\begin{eqnarray}
\psi_{\sigma}(\tau, {\bf r})  =  \phi_{0 \sigma}({\bf r})+ \phi_{\sigma}(\tau, {\bf r}),
\label{field}
\end{eqnarray}
where  $\phi_{0 \sigma}({\bf r})$ is the  condensate wave function and is independent of $\tau$.  We expand the action ${\cal S}$ in powers of the fluctuation field $\phi_{\sigma}(\tau, {\bf r})$ and obtain 
\begin{eqnarray}
 {\cal S}={\cal S}_{0}+{\cal S}_2+\cdots. 
\end{eqnarray}
The mean-field part ${\cal S}_0$ reads
\begin{eqnarray}
{\cal S}_0  =  \int {d x}\ \Big[\phi_0^{\dag}(\hat{\cal H}_s-\mu )\phi_0^{\phantom{\dag}} +\ \frac{g}{2}(n_{0\uparrow}^2+n_{0\downarrow}^2)+g_{\uparrow \downarrow}n_{0\uparrow} n_{0\downarrow}\Big], \nonumber\\
\label{S0}
\end{eqnarray}
where $\phi_0=(\phi_{0\uparrow}, \phi_{0\downarrow})^{\rm T}$ and $n_{0 \sigma}=\phi^\ast_{0\sigma}\phi_{0 \sigma}^{\phantom{\dag}}$ is the condensate density for the spin component $\sigma$. Here $x=(\tau, {\bf r})$ and $\int dx=\int_0^\beta d\tau \int d{\bf r}$. ${\cal S}_2$ denotes 
the action which is quadratic in the fluctuation field $\phi_{\sigma}(x)$. It can be expressed as 
 \begin{eqnarray}
{\cal S}_2  = \frac{1}{2}\int {d x} \Phi^\dagger(x) {\bf M} \Phi(x),
\label{S2}
\end{eqnarray}
where $\Phi=(\phi_\uparrow^{\phantom{\dag}}, \phi_\downarrow^{\phantom{\dag}}, \phi_\uparrow^*, \phi_\downarrow^*)^{\rm T}$ and the explicit form of the matrix ${\bf M}$ is given by
\begin{widetext}
\begin{eqnarray}
{\bf M}=\left(\begin{array}{cccc}\partial_\tau+{\cal{K}}_{\uparrow} & \Omega(r)+g_{\uparrow \downarrow}\phi_{0 \uparrow}^{\phantom{\dag}}\phi^{\ast}_{0 \downarrow} & g\phi^2_{0 \uparrow} & g_{\uparrow \downarrow}\phi_{0 \uparrow }\phi_{0 \downarrow} \\
\Omega(r)+g_{\uparrow \downarrow}\phi^{\ast}_{0 \uparrow}\phi_{0 \downarrow}^{\phantom{\dag}}& \partial_\tau+{\cal{K}}_{\downarrow}  & g_{\uparrow \downarrow}\phi_{0 \uparrow }\phi_{0 \downarrow} & g\phi^2_{0 \downarrow} \\
g\phi^{\ast 2}_{0 \uparrow} & g_{\uparrow \downarrow}\phi^{\ast}_{0 \uparrow }\phi^{\ast}_{0 \downarrow} & \partial_\tau+{\cal{K}}_{\uparrow} & \Omega(r)+g_{\uparrow \downarrow}\phi^{\ast}_{0 \uparrow}\phi_{0 \downarrow}^{\phantom{\dag}} \\
g_{\uparrow \downarrow}\phi^{\ast}_{0 \uparrow }\phi^{\ast}_{0 \downarrow} & g\phi^{\ast 2}_{0 \downarrow} & \Omega(r)+g_{\uparrow \downarrow}\phi_{0 \uparrow}^{\phantom{\dag}}\phi^{\ast}_{0 \downarrow} & \partial_\tau+{\cal{K}}_{\downarrow}\end{array}\right),
\end{eqnarray}
where  
\begin{eqnarray}
{\cal{K}}_{\sigma}=-\frac{\hbar^2}{2Mr}\frac{\partial}{\partial r}\left(r\frac{\partial}{\partial r}\right)+\frac{(L_z-sl\hbar)^2}{2Mr^2}+s\frac{\delta}{2}+V_{\text{ext}}({\bf r})-\mu+2gn_{0 \sigma}+g_{\uparrow \downarrow}n_{0 \bar{\sigma}}.
\end{eqnarray}
\end{widetext}
Here we define $s=+1$ for $\sigma=\uparrow$ and $s=-1$ for $\sigma=\downarrow$, and $\bar{\sigma}$ denotes the opposite spin state of $\sigma$.

The ground-state phase (zero-temperature) diagram of a BEC with SOAM coupling can be determined by minimalizing ${\cal S}_0$ in Eq.~(\ref{S0}). To this end, we express $\phi_{0\sigma}({\bf r})$ in a general form, 
\begin{eqnarray}
\begin{split}
\phi_{0\sigma}({\bf r})& =  \sum_{nm}c_{nm \sigma}R_{n,m-sl}(r)\Theta_{m}(\theta).
\label{condensate-wave}
\end{split}
\end{eqnarray}
The basis for the radial part is chosen as the eigen radial wave function of the 2D harmonic potential, which reads
\begin{eqnarray*}
R_{n,m}(r)=\frac{1}{a_\perp}\sqrt{\frac{2n!}{(n+|m|)!}}\left(\frac{r}{a_\perp}\right)^{|m|}e^{-\frac{r^2}{2a^2_\perp}}L^{|m|}_{n}\left(\frac{r^2}{a^2_\perp}\right),
\end{eqnarray*}
with $a_\perp=\sqrt{\hbar/(M\omega_\perp)}$ being the characteristic length.
The variational coefficients $c_{nm \sigma}$ is to be determined by minimizing the ground-state energy under the normalization condition   
$\sum_{nm\sigma}|c_{nm\sigma}|^2=N$, where $N$ is the total particle number. The ground-state energy reads
\begin{eqnarray}
E_G&=&  \sum_{nm\sigma}\varepsilon_{n m\sigma}|c_{nm \sigma}|^2 
\nonumber\\
&&+ \sum_m \int r dr \ \Omega(r)\left[U^*_{m \uparrow}   U_{m \downarrow}+\rm{h.c.}\right] \nonumber\\
&&+ \int d{\bf r}\left[\frac{g}{2}\left(n_{0\uparrow}^2 +n_{0\downarrow}^2 \right)+g_{\uparrow \downarrow}n_{0\uparrow}n_{0\downarrow}\right],
\label{EG}
\end{eqnarray}
with the notations $\varepsilon_{nm\sigma}=\left(2n+|m-sl |+1\right)\hbar \omega_\perp+s\delta/2$ and $U_{m\sigma}(r)=\sum_n c_{nm \sigma}R_{n,m-sl}(r)$.  

\begin{figure}[t]
\begin{center}
\includegraphics[width=0.48\textwidth]{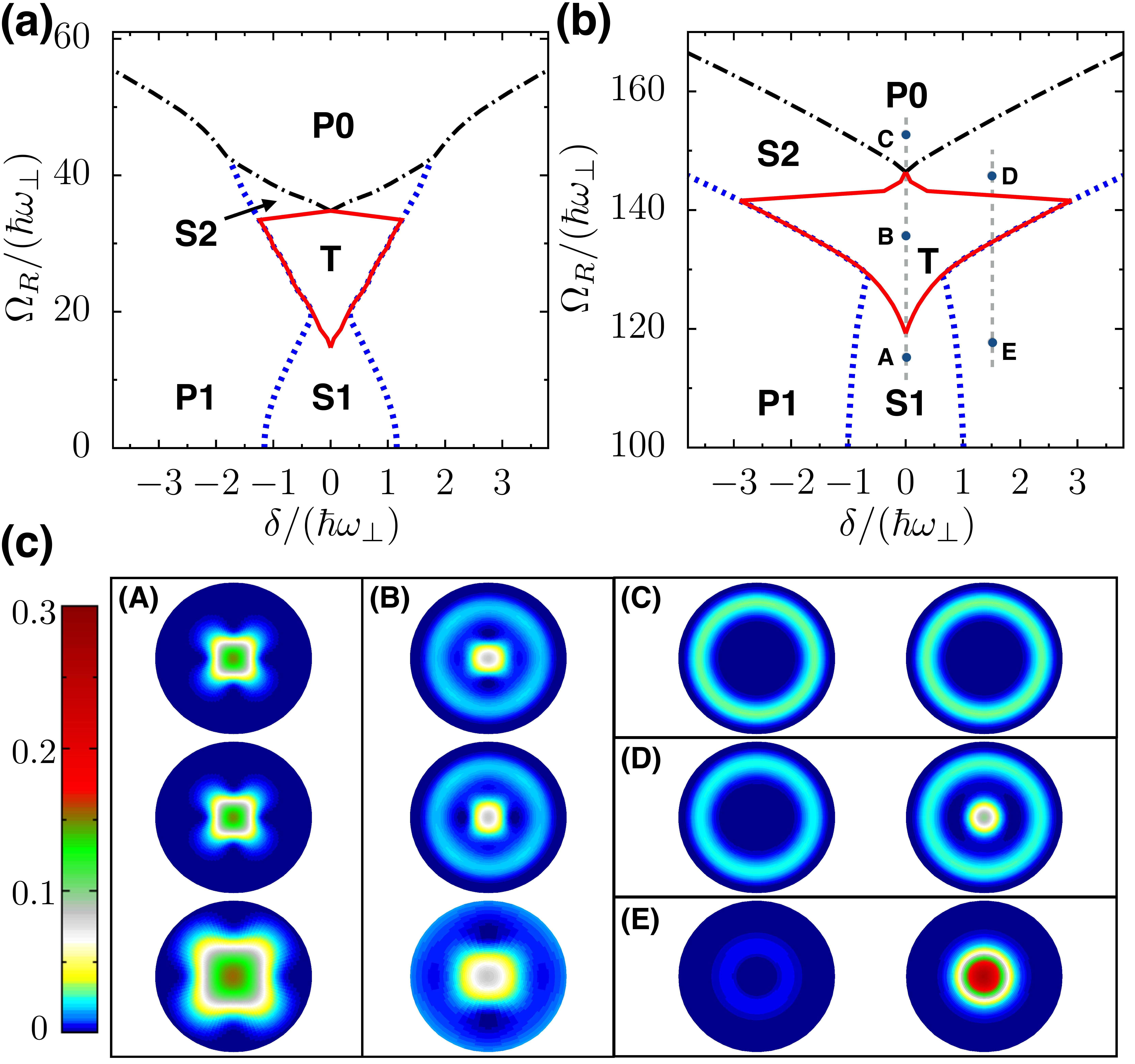}
\caption{ (a) and (b) Ground-state phase diagrams of a spinor BEC with SOAM coupling in the $\Omega_R-\delta$ plane for $l=1$ (a) and $l=2$ (b).  (c) Typical density profiles of different phases for the $l=2$ case.  The profiles shown in (A)--(E) corresponds to the points A--E in the phase diagram (b). In (A) and (B), the last plots are the zoom-in version of the $\uparrow$ component.  The interaction parameters are chosen as $a_s\simeq100a_B$ ($^{87}$Rb atoms) and $a^{\uparrow \downarrow}_{s}=50 a_B$ with $a_B$ being the Bohr radius, corresponding to $g/g_{\uparrow\downarrow}=2$. The trapping frequency along the $z$ direction is  $\omega_z=2\pi \times 200$Hz and the total particle number is $N=1000$. Other parameters are the same as in Fig. \ref{noninteraction}. }
\label{interacting-phase}
\end{center}
\end{figure}

In realistic numerical calculations, we consider a weakly interacting quasi-2D system, which can be realized by imposing a very strong harmonic confinement along the $z$ direction, with frequency $\omega_z$. In this quasi-2D system, three energy scales exist essentially. One is the characteristic energy scale along the $z$ direction, $\hbar \omega_z$.  The other two are $g N/S$ and $\hbar \omega_\perp$, which denote the interaction and single-particle energy scales, respectively. Here, $S$ is area of the 2D system. In our calculations,  we require $g N/S \lesssim  \hbar \omega_{\perp}$ and  $\hbar \omega_{\perp} \ll  \hbar \omega_{z} $, which means all atoms  are frozen at the ground state along the $z$ direction. Thus the wave function along the $z$ direction can be well approximated as the ground state of 1D harmonic potential, with its characteristic length $a_z=\sqrt{\hbar/(M \omega_z)}$. The effective 2D couplings constants in (\ref{EG}) are given by $g=\sqrt{8\pi}\hbar^2 a_s/(Ma_z)$ and $g_{\uparrow \downarrow}=\sqrt{8\pi}\hbar^2 a^{\uparrow \downarrow}_s/(Ma_z)$, where $a_{s}$ and $a^{\uparrow \downarrow}_s$ are the 3D s-wave scattering lengths for intra- and interspecies interactions, respectively. 

To study the ground-state phase diagram, we minimize the ground-state energy $E_G$ with respect to the variational parameters $c_{nm\sigma}$ numerically.  Different phases are characterized by the angular distribution 
\begin{eqnarray}
\begin{split}
P_{m}=\frac{1}{N}\sum_{n  \sigma}|c_{mn\sigma}|^2,
\label{condensate-wave}
\end{split}
\end{eqnarray}
corresponding to the occupation number for a given quantum number $m$. The ground-state phase diagrams for OAM transfer $l=1$ and $l=2$ at $g/g_{\uparrow\downarrow}\simeq 2$ are shown in 
Figs.~\ref{interacting-phase}(a) and 2(b).
Rich phases are found in the $\Omega_R-\delta$ plane, including the vortex-antivortex pair phase (P0), the half-skyrmion phase (P1),  and three different angular-stripe phases (S1, S2, and T)~\cite{phase}.

For large enough Rabi frequency $\Omega_R$,  the ground state is the so-called vortex-antivortex pair phase denoted by P0, which is characterized by the angular distribution $P_0=1$ and $P_{m\neq 0}=0$. This can be understood by the fact that ground state of the single-particle Hamiltonian is located at $m=0$.  For small Rabi frequency $\Omega_R$ and small detuning $\delta$,  the ground state is the angular-stripe phase denoted by S1, which is characterized by a double occupation of the $m=+l$ and $m=-l$ states, i.e., $P_{\pm l}\neq 0$ and $P_m$ is vanishingly small for $m\neq\pm l$.  While this angular-stripe state appears only at vanishing detuning for the 
noninteracting case, its regime in the phase diagram is broadened by the interaction effect.  With increasing detuning $\delta$, the difference between $P_{+l}$ and $P_{-l}$ becomes larger and larger. For sufficiently large detuning and small Rabi frequency, the system finally enters the half-skyrmion phase denoted by P1, characterized by the distribution $P_{+l}\neq 0$ (or $P_{- l}\neq 0$) and $P_m=0$ for all other values of $m$.  At moderate values of the Rabi frequency, two new angular-stripe phases, denoted by T and S2, appears in the phase diagram. The T phase, which appears at small detuning, is characterized by a triple occupation of the $m=+l$, $m=0$, and $m=-l$ states, i.e., $P_m$ is finite for $m=l,0,-l$ and is vanishingly small for all other values of $m$. The S2 phase is more exotic. It is characterized by a double occupation of the $m=0$ and $m=+l$ (or $m=-l$) states. For $l=1$, it exists only for small detuning and its regime in the phase diagram is tiny as shown in 
Fig.~\ref{interacting-phase}(a). However,  the situation is changed if we consider a larger OAM transfer, such as $l=2$ considered in Fig.~\ref{interacting-phase}(b). In this case, the S2 phase exists for much larger detuning, 
compared to the other two angular-stripe phases S1 and T.

In Fig.~\ref{interacting-phase}(c), we show typical density profiles of different phases for the $l=2$ case.  In the angular-stripe phases (S1, S2, and T),  the rotational symmetry is broken and thus the density profile becomes anisotropic.  However, these phases still have a C4 symmetry, i.e., the symmetry under the $\pi/4$ rotation.  The C4 symmetry is most pronounced in the S1 phase. However, in the T and S2 phases, the C4 symmetry fades and 
the rotational symmetry restores gradually.

For different quantum phases (P0, P1, S1, S2, T), the broken symmetries and the order parameters are the same in the sense of Landau theory  of phase transitions. Here we distinguish them according to their occupations 
of angular quantum number rather than their spatial symmetry such as the C4 symmetry we discussed for the $l=2$ case.  
Since the free energy (here the energy at zero temperature) goes smoothly with the parameters (detuning and Rabi frequency), we expect that these quantum phase transitions are continuous according to the Ehrenfest classification.

\begin{figure}[t]
\begin{center}
\includegraphics[width=0.48\textwidth]{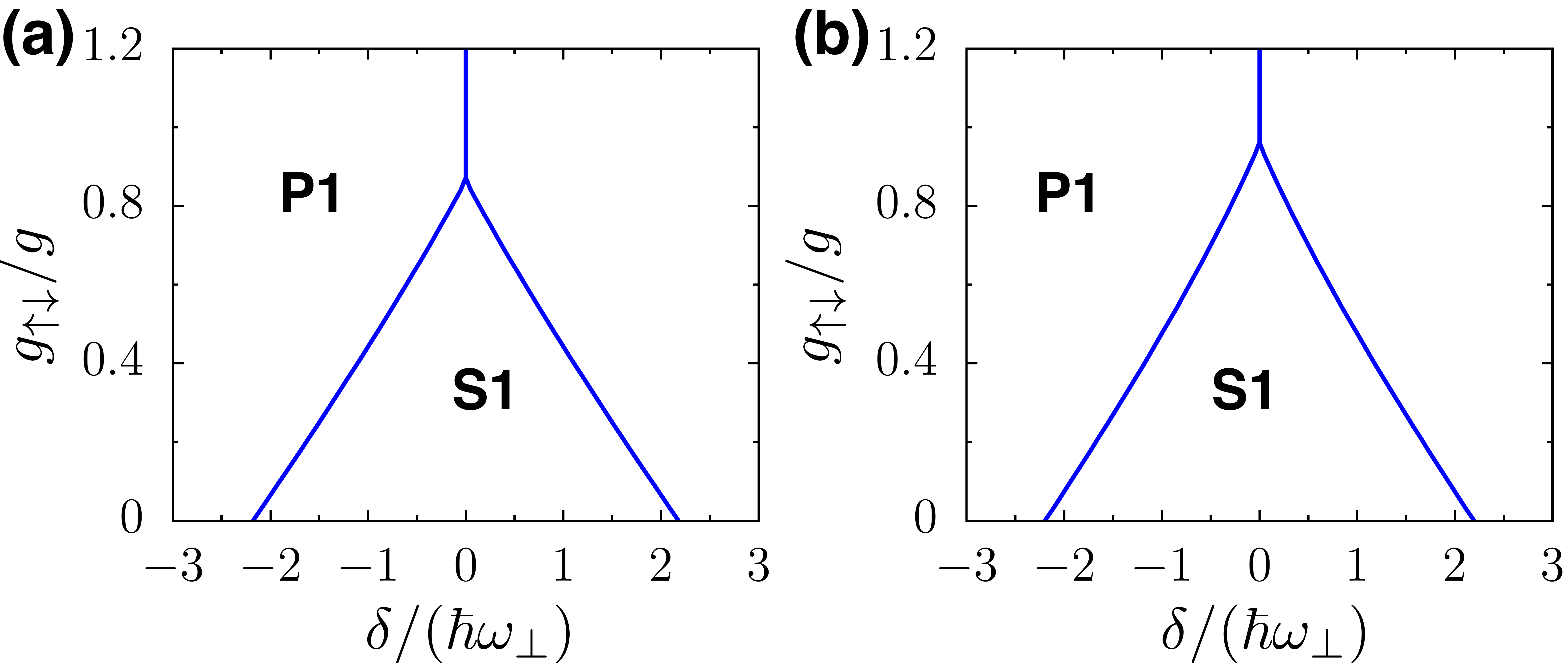}
\caption{Ground-state phase diagrams in the $g_{\uparrow\downarrow}/g-\delta$ plane: (a) $l=1, \Omega_R/(\hbar \omega_\perp)=10$ and (b) $l=2, \Omega_R/(\hbar \omega_\perp)=110$. Here the interspecies scattering length $a_s^{\uparrow\downarrow}$ is not fixed and all other parameters are the same as in Fig. \ref{interacting-phase}.}
\label{interaction-delta}
\end{center}
\end{figure}

The interaction effects on the ground-state phase diagram is shown in Fig. (\ref{interaction-delta}) \cite{TSphase}. It is evident that for large interspecies interaction, i.e.,  large values of  $g_{\uparrow \downarrow}/g$, the ground state favors the states with a single occupation of the $m$-states (P1 state for the Raman coupling we considered). While for small interspecies interaction,  i.e., small values of  $g_{\uparrow \downarrow}/g$, the ground state energetically favors the angular-stripe phase.   To reach a large regime of the angular-stripe phase,  one needs to tune
$g_{\uparrow \downarrow}/g$ as small as possible, which may be realized by using the Feshbach resonances.

\section{Excitation spectrum}\label{spectrum}
The elementary excitations of the SOAM coupled BEC, or the so-called Bogoliubov excitations, can be obtained from the action ${\cal S}_2$. 
 For the sake of simplicity, we consider the angular-stripe phase S1 and the half-skyrmion phase P1 at vanishing detuning (see Fig. (\ref{interaction-delta})).  Generalization of our calculation to other phases and nonzero detuning is straightforward but rather lengthy and complicated.  

For the S1 and P1 phases at $\delta=0$, we can restrict $m=(2\alpha-1)l$ ($\alpha\in\mathbb{Z}$) and the condensate wave function can be expressed as 
\begin{eqnarray}
\phi_{0 \sigma}({\bf r}) & =&  \sum_{n\alpha}c_{n\alpha\sigma}R_{n,(2\alpha-1-s)l}(r)\Theta_{(2\alpha-1)l}(\theta).
\label{waveII}
\end{eqnarray}
The angular-stripe phase S1 corresponds to the superposition of the $\alpha=0$ and $\alpha=1$ states (other components are vanishingly small), while the half-skyrmion phase P1 corresponds to the single occupation of the $\alpha=0$ or $\alpha=1$ state. 
The ground-state energy $E_G$ is given by Eq. (\ref{EG}) by setting $m=(2\alpha-1)l$. The chemical potential $\mu$ can be obtained from the relation  $\mu=\partial E_G/\partial N$.

To study the elementary excitations, we consider the action ${\cal S}_2$ and expand the fluctuation field $\phi_{\sigma}(\tau,{\bf r})$ using the radial and angular functions, i.e., 
\begin{eqnarray}
\phi_\sigma (\tau,{\bf r}) &=& \sum_{mn\alpha  \nu}  \phi_{n\alpha\sigma}(i\omega_\nu,m) e^{-i \omega_\nu \tau}R_{n,(2\alpha-1-s)l+m}(r)\nonumber\\
&&\times\ \Theta_{(2\alpha-1)l+m}(\theta),
\label{delta-phi}
\end{eqnarray}
where $\omega_\nu=2\pi\nu/\beta$ ($\nu\in\mathbb{Z}$) is the boson Matsubara frequency and $m$ now denotes the angular quantum number of the excitation. The summation over $\alpha$ runs from $-\infty$ to $\infty$. Substituting this expression into ${\cal S}_2$, we obtain
\begin{eqnarray}\label{effective-S2}
{\cal S}_2=\frac{\beta}{2}\sum_{m \nu}\sum_{nn^\prime\alpha\alpha^\prime}\Phi_{n\alpha}^\dagger(i\omega_\nu,m) (G^{-1})_{n\alpha,n^\prime\alpha^\prime}^{\phantom{\dag}}
\Phi_{n^\prime\alpha^\prime}^{\phantom{\dag}}(i\omega_\nu,m),\nonumber\\
\end{eqnarray}
where the notation $\Phi_{n\alpha}$ is defined as 
\begin{eqnarray}
\begin{split}
\Phi_{n\alpha}(i\omega_\nu,m) =  
\left(\begin{array}{c}\phi_{n \alpha\uparrow}^{\phantom{\dag}}(i\omega_\nu,m) \\  \phi_{n \alpha\downarrow}^{\phantom{\dag}}(i\omega_\nu,m)  \\ \phi_{n\alpha\uparrow}^*(-i\omega_\nu,-m)
\\ \phi_{n\alpha\downarrow}^*(-i\omega_\nu,-m) \end{array}\right).
\end{split}
\end{eqnarray}

The quantity $G^{-1}(i\omega_\nu,m)$, which is also a matrix in the spaces spanned by the quantum numbers $n$ and $\alpha$, characterizes the propagation of the excitation.  The Green's function $G^{-1}$ can be expressed as  
$G^{-1}=G^{-1}_0+G^{-1}_1+G^{-1}_{2}$, 
where $G^{-1}_{0}$,  $G^{-1}_1$  are given by

\begin{eqnarray}
\left(G^{-1}_0\right)_{n\alpha,n'\alpha'} &=&\left (\begin{array}{cc}- i\omega_\nu \mathbb{I}+\mathbb{X}^{n\alpha}_{m} & 0 \\0 & i \omega_\nu \mathbb{I}+ \mathbb{X}^{n\alpha}_{-m}\end{array}\right)\delta_{nn',\alpha\alpha'},  \nonumber\\
\left(G^{-1}_{1}\right)_{n\alpha,n'\alpha'} &=& \left (\begin{array}{cc}\mathbb{Y}^{n'\alpha'}_{m, n\alpha} & 0  \\ 0 &  \mathbb{Y}^{n'\alpha'}_{-m, n\alpha}\end{array}\right)\delta_{\alpha\alpha'},
\end{eqnarray}
where $\mathbb{I}$ is the $2\times 2$ identity matrix and
\begin{eqnarray}
\mathbb{X}^{n \alpha}_{m} &=&\left (\begin{array}{cc}\xi_{n\alpha \uparrow}(m) & 0 \\0 & \xi_{n\alpha \downarrow}(m) \end{array}\right),\nonumber\\
 \mathbb{Y}^{n'\alpha'}_{m, n\alpha} &=&\left (\begin{array}{cc}0 & \Omega_m(n\alpha, n'\alpha') \\ \Omega_m(n' \alpha', n\alpha) & 0\end{array}\right),
\end{eqnarray}
with  $\xi_{n \alpha \sigma}(m)=(2n+|(2\alpha-1)l+m-sl|+1)\hbar \omega-\mu$ and 
\begin{widetext}
\begin{eqnarray}
\Omega_m(n\alpha, n'\alpha') =  \int r dr R_{n,(2\alpha-1)l+m-l}(r)\Omega(r) R_{n',(2\alpha'-1)l+m+l}(r). 
\end{eqnarray}
$G^{-1}_2$ is given by
\begin{eqnarray}
\left(G^{-1}_2\right)_{n\alpha, n'\alpha'}&=&
\left (\begin{array}{cccc}A_{m\uparrow}(n\alpha,n'\alpha') & D_m(n\alpha, n'\alpha') & B_{m\uparrow}(n\alpha, n'\alpha') & C_m(n\alpha, n'\alpha') \\
D^{\ast}_m(n'\alpha',n\alpha) & {A_{m\downarrow}}(n\alpha,n'\alpha') & C_{-m}(n'\alpha',n\alpha) & B_{m\downarrow}(n\alpha,n'\alpha') \\
B^{\ast}_{-m \uparrow}(n\alpha,n'\alpha') & C^{\ast}_{-m}(n\alpha,n'\alpha') &A_{-m\uparrow}(n\alpha,n'\alpha') & D^{\ast}_{-m}(n\alpha,n'\alpha') \\
C^{\ast}_m(n'\alpha',n\alpha) & B^{\ast}_{-m\downarrow}(n\alpha, n'\alpha') & D_{-m}(n'\alpha',n\alpha) & A_{-m \downarrow}(n\alpha,n'\alpha')\end{array}\right),
\end{eqnarray}
where $A_{m\sigma}(n\alpha, n'\alpha')$, $B_{m\sigma}(n\alpha, n'\alpha')$, $C_{m}(n\alpha, n'\alpha')$, $D_{m}(n\alpha, n'\alpha')$ are defined as
\begin{eqnarray}
A_{m \sigma}(n\alpha, n'\alpha') & =&  \frac{1}{2\pi}\sum_{\alpha_1 \alpha_2}\int r dr R_{n,(2\alpha-1)l+m-sl}(r)R_{n',(2\alpha'-1)l+m-sl}(r)\left(2gU^{\ast}_{\alpha_1 \sigma}U_{\alpha_2 \sigma}^{\phantom{\dag}}
+g_{\uparrow \downarrow}U ^{\ast}_{\alpha_1 \bar{\sigma}}U_{\alpha_2 \bar{\sigma}}^{\phantom{\dag}}\right)\delta_{\alpha_1+\alpha,\alpha_2+\alpha'}, \nonumber\\ 
B_{m \sigma}(n\alpha, n'\alpha') & = & \frac{g}{2\pi}\sum_{\alpha_1 \alpha_2}\int r dr R_{n,(2\alpha-1)l+m-sl}(r)R_{n',(2\alpha'-1)l-m-sl}(r)U_{\alpha_1 \sigma}U_{\alpha_2 \sigma}\delta_{\alpha_1+\alpha_2,\alpha+\alpha'}, \nonumber\\
C_m(n\alpha, n'\alpha') & =&  \frac{g_{\uparrow \downarrow}}{2\pi}\sum_{\alpha_1 \alpha_2}\int r dr R_{n,(2\alpha-1)l+m-l}(r)R_{n',(2\alpha'-1)l-m+l}(r)U_{\alpha_1\uparrow}U_{\alpha_2 \downarrow}\delta_{\alpha_1+\alpha_2,\alpha+\alpha'}, \nonumber\\
D_m(n\alpha, n'\alpha') & = & \frac{g_{\uparrow \downarrow}}{2\pi}\sum_{\alpha_1 \alpha_2}\int r dr R_{n,(2\alpha-1)l+m-l}(r)R_{n',(2\alpha'-1)l+m+l}(r)U_{\alpha_1 \uparrow}^{\phantom{\dag}}U^{\ast}_{\alpha_2 \downarrow}\delta_{\alpha_1+\alpha',\alpha_2+\alpha}.
\end{eqnarray}
\end{widetext}
Here $U_{\alpha\sigma}(r)=\sum_n c_{n\alpha \sigma}R_{n,(2\alpha-1)l-sl}(r)$. From the above definitions, we have $A_{m\sigma}(n\alpha, n'\alpha')=A_{m\sigma}(n'\alpha',n\alpha)$ and $B_{m\sigma}(n\alpha,n'\alpha')=B_{m\sigma}(n'\alpha',n\alpha)$. 

\begin{figure}[t]
\begin{center}
\includegraphics[width=0.48\textwidth]{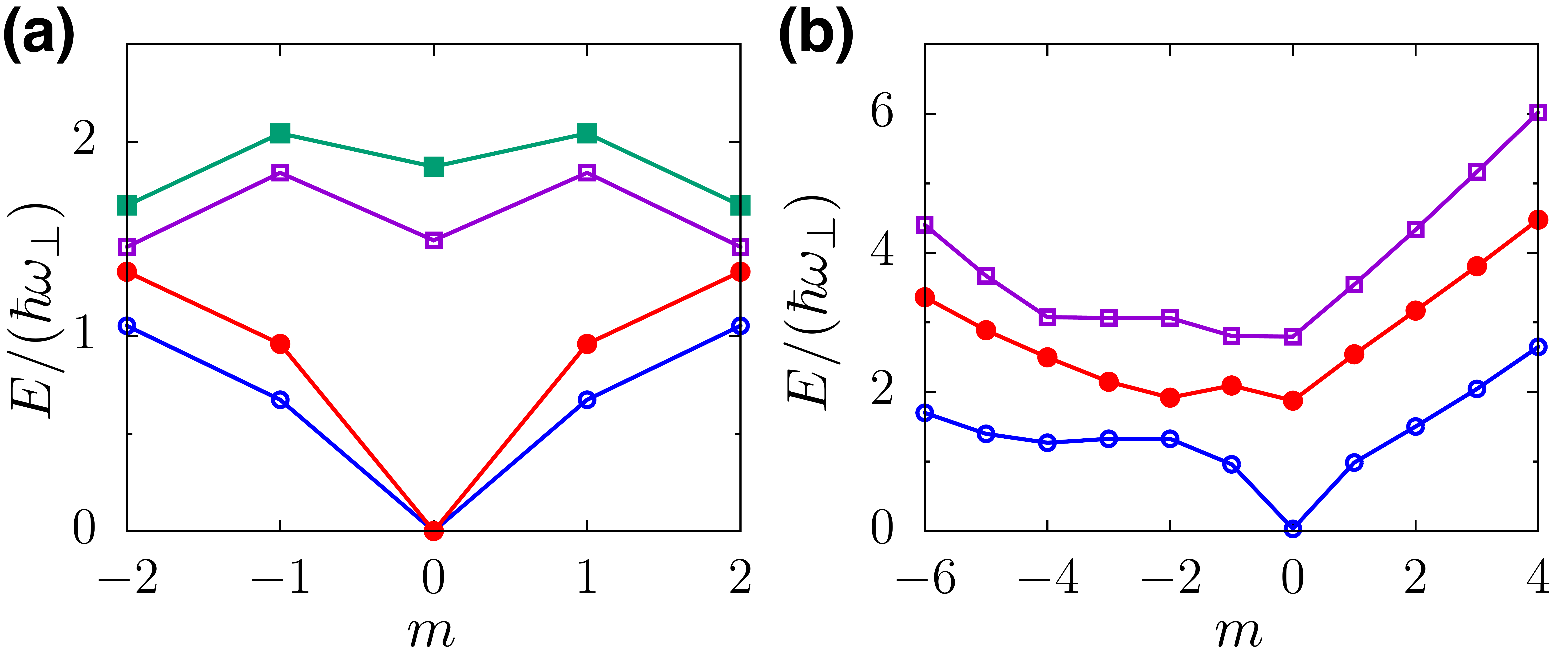}
\caption{ Excitation spectra of an interacting BEC with SOAM coupling in the angular-stripe phase S1 (a) and in the half-skyrmion phase P1 (b). The interaction parameters are chosen as $g_{\uparrow \downarrow}/g=1/2$ (a)  and $g_{\uparrow \downarrow}/g=2$ (b). The detuning $\delta$ is zero, $\Omega_R/(\hbar \omega_{\perp})=30$,  and all other parameters are the same as in Fig. \ref{interacting-phase}.}
\label{excitation}
\end{center}
\end{figure}

The excitation spectrum $E(m)$ can be obtained by
\begin{eqnarray}
\det \left[G^{-1}(E,m)\right]=0
\end{eqnarray}
after the analytical continuation $i\omega_\nu\rightarrow E+i0^+$, where the determinant is taken also in the spaces spanned by the quantum numbers $n$ and $\alpha$.
The excitation spectra in the angular-stripe phase S1 and the half-skyrmion phase P1 for $l=2$ are shown in Figs.~\ref{excitation}(a) and 4(b), respectively.  In both phases, there exists a gapless band,
i.e., the Goldstone mode corresponding to the broken U$(1)$ symmetry. In the angular-stripe phase S1 [Fig. \ref{excitation}(a)],  we also find an additional gapless band corresponding
to the broken rotational symmetry. In the half-skyrmion phase P1 [Fig. \ref{excitation}(b)],  the excitation spectrum exhibits roton-like structure around $m=-4$.  We expect that some exotic features,  
for example, two zero modes of angular-stripe phase, softening of roton in the half-skyrmion phase can be observed in feature experiments.

\section{summary}\label{summary}
In summary,  we have studied the ground-state phase diagram and excitation spectrum of an interacting spinor BEC with SOAM coupling realized in recent experiments. For a large OAM difference $l=2$, we predict an unconventional angular-stripe phase which can exist at large detuning. The excitation spectra in the angular-stripe phase and the half-skyrmion phase have distinct features. These findings would be helpful to identify the interesting angular-stripe phase in future experiments.

\begin{acknowledgements} We acknowledge fruitful discussions with Shi-Guo Peng and Xiao-Long Chen. This work was supported by the National Key Research and Development Program of China (Grant No. 2018YFA0306503) and National Natural Science Foundation of China (Grant Nos. 11775123 and 11890712).
\end{acknowledgements}


\bibliographystyle{apsrev4-1}
\bibliography{myref}

\begin{thebibliography}{33}
\bibitem [{exp()}]{explain01}%
  \BibitemOpen
  \href@noop {} {}\bibinfo {note} {In principle, $l_1$ and $l_2$ can be
  arbitrary integers and hence $l$ can be either integer or half integer. In
  this work, we consider the case $l_1=0$ and $l_2=-2l$ with $l$ integer, which
  has been experimentally realized. For other choices of $l_1$ and $l_2$ the
  ground-state phase diagram will not be qualitatively modified (even if $l$ is
  half integer) since the ground-state degeneracy of single-particle
  Hamiltonian will not change essentially.}\BibitemShut {Stop}%
\bibitem [{pha()}]{phase}%
  \BibitemOpen
  \href@noop {} {}\bibinfo {note} {For $l=2$, the P0 and P1 phases correspond
  to the phases with quantum numbers $(-2,2)$ and $(0,4)$ or $(-4,0)$,
  respectively. See Fig. 3 of Ref. \cite{qu-15}. Another phase with quantum
  number $(-3,1)$ or $(-1,3)$ does not appear in our phase
  diagram.}\BibitemShut {Stop}%
\bibitem [{TSp()}]{TSphase}%
  \BibitemOpen
  \href@noop {} {}\bibinfo {note} {The absence of the T and S2 phases in
  Fig.(\ref{interaction-delta}) is because we choose a relatively small Rabi
  frequency $\Omega_R$. For larger values of $\Omega_R$, the S1 phase may be
  replaced by the T phase and the S2 phase may also appear between T and P1
  phases.}\BibitemShut {Stop}%
\end{thebibliography}

\end{document}